# Nitrogen Doping and Infusion in SRF Cavities: A Review

Pashupati Dhakal[*]

Thomas Jefferson National Accelerator Facility, Newport News, VA 23606, USA

*Abstract*

Advances in SRF technology over the last 40 years allowed achieving accelerating gradients ~ 50 MV/m corresponding to peak surface magnetic fields close to the theoretical limit of niobium. However, the quality factor decreases significantly with increasing accelerating gradient. This decrease is expected since increasing the rf field increases the number of quasiparticles and therefore the rf losses. Recently, a new phenomenon of increase in quality factor with the accelerating gradient has been observed when SRF cavities are doped with certain non-magnetic impurities. In particular, the diffusion of nitrogen into the niobium cavities' inner surface has been successfully implemented into the commercialization of SRF technology. The quest is still ongoing towards process development to achieve high accelerating gradient SRF structures with high-quality factors for future high energy accelerators. Here, we present the review of the research on process development to this date via nitrogen diffusion, material analysis, and present the summary of current theoretical understanding behind high $Q_0$ SRF cavities.

## I. Introduction

Superconducting radiofrequency (SRF) technology is being used not only for the basic fundamental nuclear physics research but also for applications that have benefited society [1]. Modern particle accelerating machines rely on SRF technology, few examples of current and future projects are continuous wave (CW) free-electron lasers, x-ray laser oscillators, light sources, electrons and ions colliders, accelerator-driven systems for medical isotope production and nuclear waste transmutation. SRF technology which is based on the superconducting hollow structures (cavities) with high duty factor provides the required accelerating gradient to accelerate the charged particles close to the velocity of light.

The superiority of superconducting cavities over the normal metal cavities is their ability to store a large amount of energy with much lower power dissipation. The performance of the SRF cavities is measured in terms of the quality factor $Q_0=G/R_s$, where $G$ is the geometric factor which depends on the cavity geometry and $R_s$ is the surface resistance, as a function of accelerating gradient, $E_{acc}$. For typical cavities of resonating frequency 0.5–3 GHz operating at a temperature ~2.0 K the quality factor is in the $10^{10}$–$10^{11}$ range. In the last four decades, much research work has been focused on improving the accelerating gradient and quality factor of these SRF cavities.

---
[*] Email: dhakal@jlab.org

Considerable effort has gone into pushing the quality factor to higher values as well. Higher $Q_0$ for the reduction of cryogenic losses, requiring liquid helium cryogenic plants with lower cooling capacity, and higher $E_{acc}$ to achieve higher beam energy and smaller footprint are desired. Some of $E_{acc}$ and $Q_0$ requirements for some current and future SRF based accelerator is shown in Fig. 1. Typically, different accelerators have different specifications for the gradient and the quality factor. Machines such as the International Linear Collider and the European X-ray Free Electron Laser XFEL are designed to operate in pulsed mode whereas the upgrade of the Linear Coherent Light Source (LCLS-II), accelerator-drives systems (ADS), and Energy Recovery Linacs are designed to operate in CW mode [2,3]. For CW applications, the minimization of the overall AC plug power limits the optimal gradient to ≤20 MV/m [4]. Continuous research and development has already pushed the achievable accelerating gradient towards the fundamental limit of Nb, although not in a reproducible way [5].

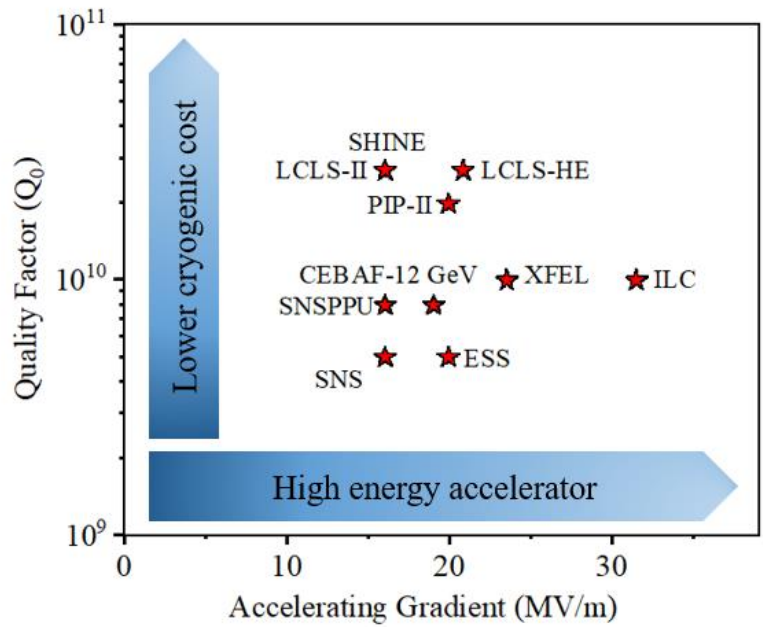

**Figure 1.** Schematic showing the requirements for cavity performances for different current and future SRF based accelerators

The current and future accelerators would highly benefit from cavities with higher quality factor and accelerating gradient. A high quality factor can be achieved by minimizing the surface resistance of SRF cavities. The surface resistance in superconducting materials is the sum of the temperature-independent residual resistance ($R_{res}$) and temperature-dependent Bardeen-Cooper-Schrieffer (BCS) resistance ($R_{BCS}$) [6,7]. The sources of the $R_{res}$ are trapped magnetic flux during the cavity cool down, impurities, hydrides and oxides, imperfections, and surface contamination. The BCS surface resistance results from the interaction between the rf electric field within the penetration depth and thermally activated quasi-particles in the superconductor. The BCS surface resistance depends on superconducting material parameters which may vary strongly due to the presence of the impurities within the rf penetration depth.

In the past decade, the development of high-quality factor SRF cavities via material diffusion in a thin layer at the inner surface of the cavities has been achieved. It was first realized by serendipitous titanium doping during annealing at high temperature (~1400 °C) without any post-annealing chemistry [8,9] and later by serendipitous nitrogen doping at 800 °C, followed by electropolishing [10]. The impurity diffusion process not only resulted in an increase in quality factor at low rf field but also an increase in quality factor with increasing accelerating gradient, contrary to the previously observed *Q*-slope. However, it was also found that the maximum accelerating gradient was reduced on average by ~30 % as a result of the doping process.

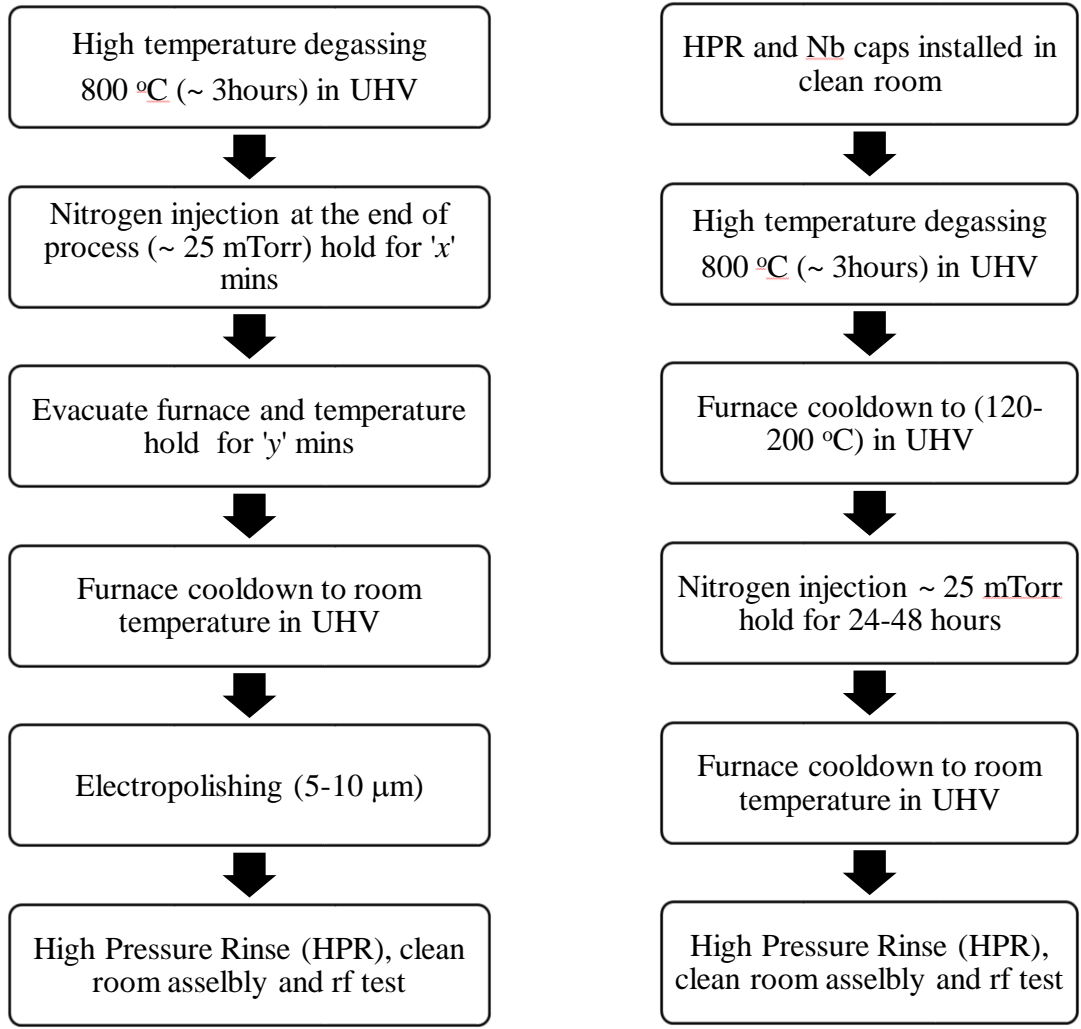

**Figure 2.** Final process flow for nitrogen doping (left) and nitrogen infusion (right).

Most recently, efforts have been made to preserve high accelerating gradients while also increasing the quality factor of SRF cavities. In these new nitrogen "infusion" cavity processing recipes, cavities were heat-treated at high temperature (>800 °C), then the furnace temperature is reduced to 120–200 °C, and nitrogen was introduced into the furnace at a partial pressure of ~25

mTorr for several hours. This process has shown an improvement in $Q_0$ over the baseline measurements, without the need for post-annealing chemical removal of material from the inner surface [11,12,13]. Even though the diffusion of the nitrogen into the bulk of the SRF cavity is limited in-depth at these low temperatures, the introduction of nitrogen is sufficient to modify the cavity surface within the rf penetration depth as seen from rf results, which are similar to those previously reported for high-temperature nitrogen-doped cavities. Furthermore, while post-doping electropolishing is required to remove coarse lossy nitrides from the surface, no further processing is required for the low-temperature infusion recipe showing a clear benefit in reducing processing steps as well as keeping higher gradient with high $Q_0$ values. Figure 2 shows the typical final cavity processing steps during the nitrogen doping and infusion process. The nitrogen doping recipe with x= 2 mins and y=6 mins was rapidly implemented into the industrialization of the technology as the processing recipe of SRF cavities for LCLS-II cryomodules production.

The historical development of SRF technology over the last four decades can be found in Ref. [14] and several conferences and workshop proceedings [15]. This paper is organized as follows. In section II, we present the invention of the nitrogen doping recipe. In section III, we present the recipe developments related to furnace requirements, nitrogen diffusion time, and post doping electropolishing. In section IV, we comment on the industrialization efforts related to the LCLS-II cavity production. In section V, we present the recent development on recipe modification towards the high gradient, high quality factor SRF cavities. In section VI, we present the sample studies to understand the mechanism behind the improvement in SRF cavity performances. In section VII, we present some theoretical models that explain the high $Q$ and high gradient cavity performance. Section VIII presents the summary and future outlooks.

## II. The invention of Nitrogen Doping

Over the last several decades, several attempts were made to increase the quality factor of SRF cavities. The Nb surface is covered by a few nanometer-thick oxide structure and higher concentration of interstitials at the metal-oxide interface. Work was done to passivate the Nb surface with a less-defective oxide [16] or with a thin nitride layer nitrides [17,18] to prevent a high concentration of the surface hydrogen leading to the formation of Nb-H phase which can result in a degraded cavity performance [19].The surface oxidation experiments were carried out at high temperature and showed the enhanced $Q_0$ [16]. In 2012, it was serendipitously discovered that the diffusion of a small amount of titanium during the high-temperature heat treatment of a Nb cavity was found to be related to to the increase in quality factor with the accelerating gradient, as shown in Fig. 3(a) [8,9].

An exploration similar to that of Ref. [18] was carried out at Fermi Lab in 2013 in an attempt to form niobium nitride on the inner surface of the SRF cavity by reacting bulk niobium cavities with nitrogen in a high-temperature UHV furnace [10]. The process resulted in the formation of lossy, metallic NbN phases on the surface, but it was discovered that by successfully removing thin layers of material by electropolishing an increase in quality factor with accelerating gradient, as that shown in Fig. 3(b), was obtained was obtained after removing ~5-7 μm of material. The $Q_0(E_{acc})$ curves in Fig. 3 show similar anti-Q-slope indicating that the interstitial diffusion of the non-magnetic impurities (Ti and N) indeed enhance the quality factor by decreasing the BCS surface resistance as the accelerating gradient increases up to 10-15 MV/m. The advantage of nitrogen doping over the Ti doping is that nitrogen being gaseous, it is easier to control the partial pressure in a high-temperature environment and nitrogen diffusion takes place at a lower temperature compared to titanium, which required a temperature higher than 1250 °C [8].

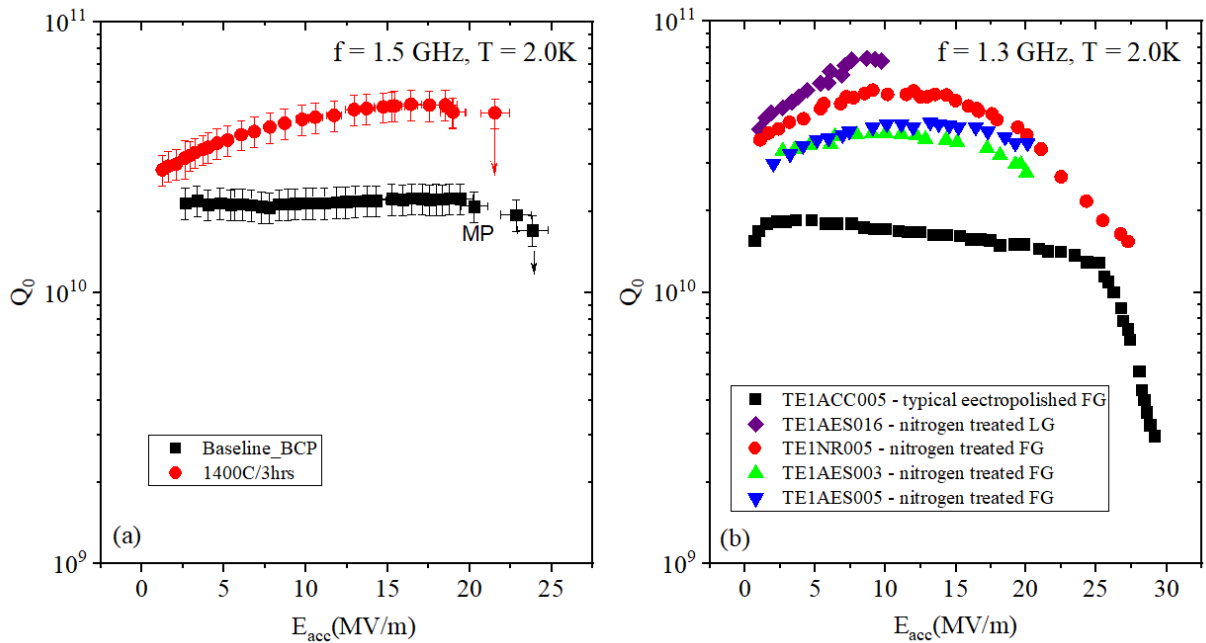

**Figure 3**. The first demonstration of an increase in quality factor with rf field with (a) titanium doping on 1.5 GHz single-cell cavity made from large grain Nb at 2.0 K [8] and (b) nitrogen doping followed by electropolishing on 1.3 GHz cavities at 2.0K [10].

## III. Nitrogen Doping Recipe Development

After the demonstration of an exceptionally high quality factor on nitrogen-doped SRF single-cell cavities, the focus quickly shifted to the application of nitrogen-doped cavities in multi-cell cavities and eventually in a superconducting linear accelerator. Around the same time, the LCLS-II project quickly adopted nitrogen-doping as part of the treatment processes for the SRF cavities to be installed in cryomodules [20]. The LCLS-II baseline design required 1.3 GHz 9-cell TESLA shaped cavities [21] with an average intrinsic quality factor $Q_0 \sim 2.7 \times 10^{10}$ at an accelerating gradient of 16 MV/m at 2.0 K [22]. Joint R&D efforts among different laboratories towards achieving the high quality factor were started with aiming towards the development of the cavity processing technique that can be commercially produced in large numbers [23]. Three partner labs to LCLS-II projects: Fermi Lab, Jefferson Lab, and Cornell University rapidly developed separate recipes with nitrogen doping on single-cell cavities and they were successfully applied to 9-cell cavities [24,25,26,27].

### A. Doping Heat Treatment

Generally, most of the SRF facilities are equipped with a vacuum furnace capable of reaching a temperature of 1250 °C with resistive heaters. The SRF facilities at Fermi Lab [28], Jefferson Lab [29] and Cornell University [30] used a furnace manufactured by TM Vacuum Products, Inc. [31]. These furnaces are equipped with dry roughing pumps, cryopumps, residual gas analyzers, vacuum gauges, and gas delivery systems. The high purity nitrogen (> 99.9999%) was used during the doping process with the flow controlled by a mass flow controller. Typically, the temperature of the furnace is ramped up to the desired temperature (800-1000 °C) at the rate of 3-5 °C/min and hold for few hours (~3 hours), mainly for hydrogen degassing. At the end of the degassing cycle, high purity nitrogen gas is injected for some time (2-30 min), also called soaking time, with the specified partial pressure of nitrogen (~25 mTorr). During the early stage of the recipe development, the gas was turned off, the furnace was evacuated and the furnace was naturally cool down to the room temperature. Later, the recipe was modified in such a way that after the soaking time, the furnace was evacuated and the cavity was further annealed before cooling down to room temperature. A typical nitrogen doping temperature and pressure profile is shown in Fig. 4. Generally, the high Q was observed with all of the doping recipe, however, a research group at KEK initially was not able to reproduce the doping results [32] which they eventually did after improving the furnace pumping system [33]. Surface contamination for example by hydrocarbons and residual metallic particles in furnaces was found to be detrimental to the cavity's rf performance.

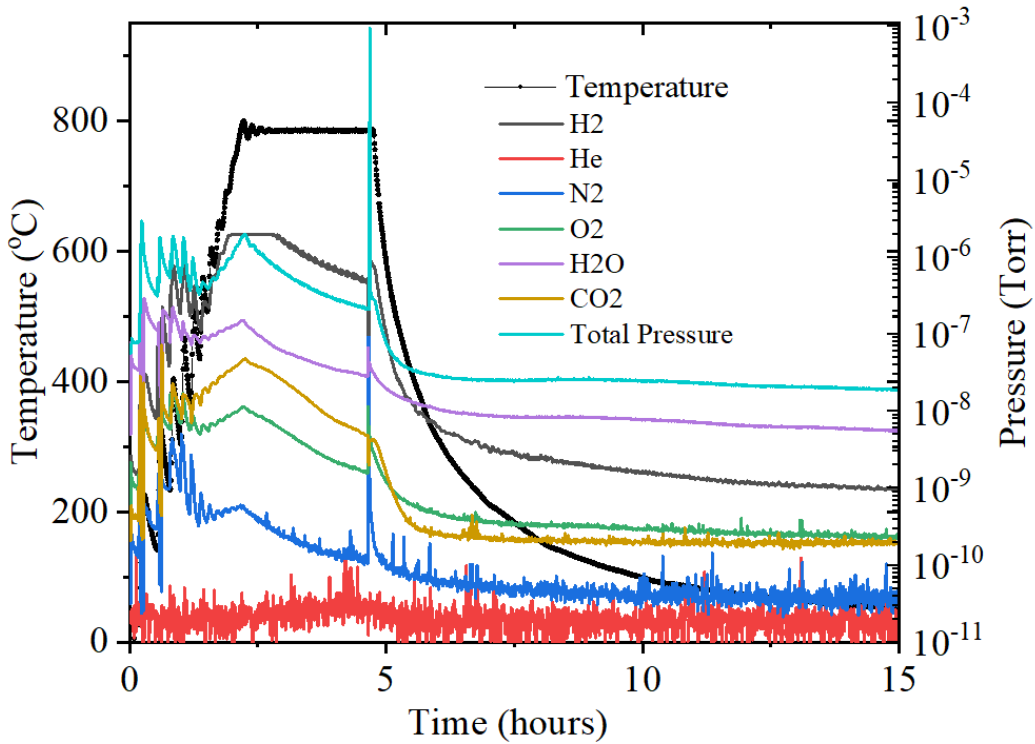

**Figure. 4** Typical temperature and pressure profile with residual gas analyzer during the nitrogen doping at Jefferson Lab. The nitrogen was injected into the furnace at 800 °C for 2 mins and the cavity was further annealed in UHV for 6 mins before cooldown to room temperature.

B. Nitrogen Diffusion

Thermal diffusion of nitrogen into Nb is a method that had been used in the past to produce NbN films [34,35,36], with higher superconducting transition temperature than Nb. Earlier attempts to produce the nitride phase for surface passivation in SRF cavities didn't result in the Q-rise phenomenon, however, the overall increase in quality factor was observed at all accelerating gradients when the SRF cavities were heat-treated in a nitrogen environment at temperature ~400 °C with the partial pressure of ~ $10^{-5}$ mTorr [18]. The study suggested that the temperature and partial pressure of nitrogen were insufficient to diffuse into the niobium or to formation of any nitride phase.

Historically, the Nb-N system is interesting because of the superconducting behavior of the δ-NbN$_{1-x}$ phase [37]. Several multi-phase regions exist in the equilibrium phase diagram of Nb-N depending on N: Nb ratio as shown in Fig. 5. In the low N concentration regime below 10 at % of N, the main phase is that of nitrogen as an interstitial in Nb, and as the concentration increases there is form a two-phase mixture of α-Nb(N), where N is an interstitial and a stoichiometric β-Nb$_2$N phase [38]. From the phase diagram, the phase of the nitride on niobium is determined by the concentration of the nitrogen and temperature of the niobium substrate [39,40]. The apparent activation energy during the nitrogen diffusion decreases with increasing temperature up to about 650 °C. Then it stays constant up to 1300 °C, above which it increases to a new value of 50 kcal/mol [41]. Most of the diffusion data available in the literature is done at a higher temperature with atmospheric nitrogen pressure on sample coupons as shown in Fig. 6.

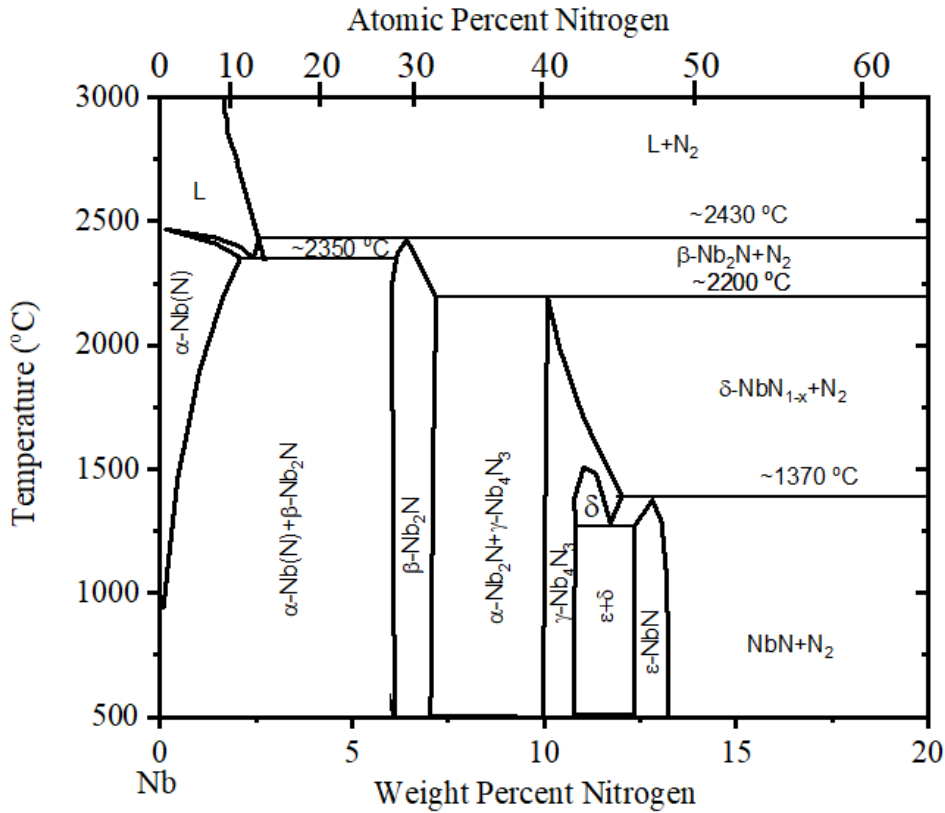

**Figure 5.** Nb-N Phase diagram, adapted from Ref. [37]

With the first demonstration of N-doping in SRF cavities, more attention towards the diffusion mechanism of nitrogen into niobium was given. The diffusion of the nitrogen in niobium usually starts with the formation of the stoichiometric nitride phase on the surface and diffusion of the nitrogen in the bulk takes place. Depending on the temperature and amount of nitrogen being introduced on the surface of niobium, the diffusion depth of nitrogen can be few to a hundred micrometers [42, 43, 44, 45]. The nitride features are visible on the surface of niobium and those are identified as $Nb_2N$ by TEM analysis [46] as well as XPS measurements [47, 48] consistent with earlier observations of sub nitride having a composition of hexagonal $Nb_2N$ [49]. Tetragonal, a face-centered cubic, and two hexagonal nitrides having compositions of $NbN_{0.75}$ to $NbN$ were also observed in earlier studies when nitrogen diffusion was done at atmospheric pressure and higher temperature [50]. At the temperature of interest for SRF cavities, a very small amount of nitrogen will diffuse to the bulk of the Nb, and most forms nitride layers on the surface starting at a temperature as low as 400 °C [51].

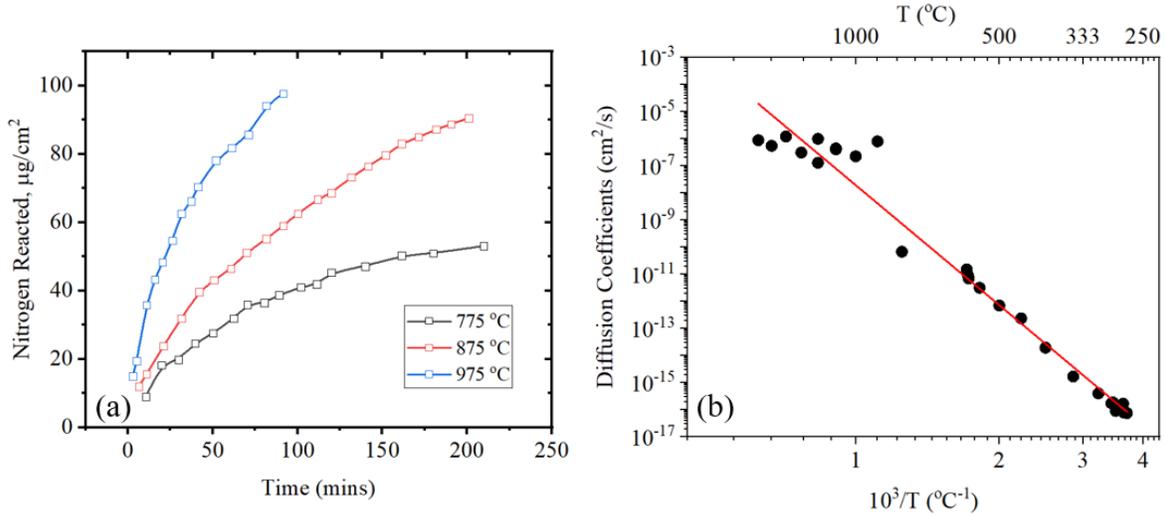

**Figure 6.** (a) Reaction rate of nitrogen in atmospheric pressure [49]. (b) Nitrogen diffusion coefficient as a function from temperature [39].

Very limited work has been done on the dependence of nitrogen diffusion on crystal orientation in high-purity SRF niobium [41, 52]. Since niobium is a body-centered cubic structure, it should not exhibit a diffusion profile based on the crystal orientation. The doping study performed on cavities made from large grain niobium also showed the improved $Q$ as good as in cavities made from fine grain niobium, if not better [10, 44, 53, 54]. In SRF cavities with large surface areas, the diffusion of nitrogen along the grain boundaries and dislocations networks is possible. No performance enhancement was observed on a cavity made from low purity niobium, probably due to the higher concentration of impurities preventing the optimal diffusion of nitrogen in the bulk [44]. However, more systematic studies are still needed to understand the mechanism of diffusion of nitrogen in niobium to the temperature, partial pressure of nitrogen, duration of nitrogen exposure, and the metallurgical state of niobium.

### C. Post-Doping Electropolishing

One of the required steps after the nitrogen doping to achieve high quality factor is controlled electropolishing of the inner surface of the cavity. A recent study on a single cell cavity also showed an encouraging result when the post doping material removal was done by buffered chemical polishing (BCP) [55]. Historically, the EP process is applied to achieve a smoother cavity surface and it was found to be superior to the BCP process. The standard EP acid mixture contains nine-volume parts of sulphuric acid $H_2SO_4$ (96%) and 1 part of hydrofluoric acid HF (48%). EP is a surface finishing process whereby the anodization of Nb by $H_2SO_4$ forces the growth of $Nb_2O_5$ and $F^-$ dissolves $Nb_2O_5$. A lot of progress has been made in optimizing the electropolishing technique over the years [[56, 57, 58,59,60]. The typical horizontal electropolishing systems consist of the cathode-anode set up with the cavity being anode and an aluminum tube being the cathode inserted along the cavity axis as shown in Fig. 7. The I-V characteristics, the temperature of the acid mixture, the temperature of the cavity surface play a significant role in etching or polishing of the niobium surface. The details on process development can be found in Ref. 61.

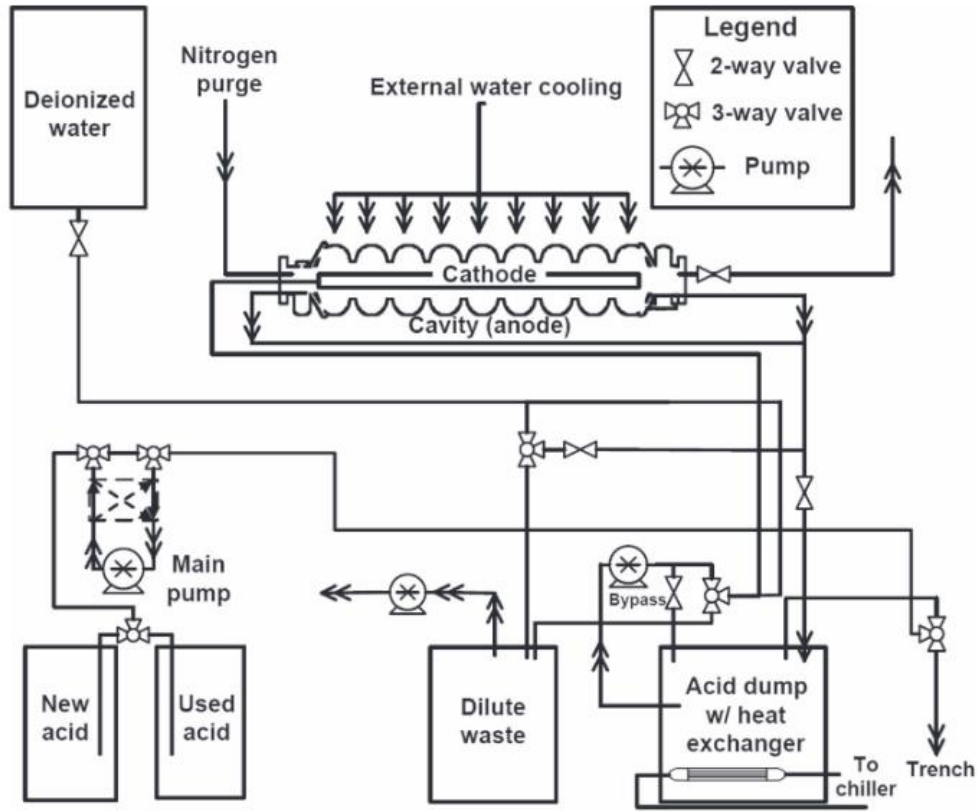

**Figure 7.** Schematic of a cavity electropolishing system [56].

The effect of the amount of material removal on the performance of nitrogen-doped cavities was studied in details [23-30]. It was found that the accelerating gradient increases while decreasing the overall quality factors as more material was removed from the cavity's inner surface, especially in heavily doped SRF cavities, as shown in Fig. 8. This confirms that some optimal nitrogen concentration is needed for high $Q_0$, however, the gradient is limited to the lower values. Questions persist on the relationship between material removal and $Q_0(E_{acc})$. How much material removal is necessary for optimal performance? It is believed that the early cavity quenches are due to the large segregation of nitrogen on a localized site, driving the region into a normal conducting state. Further removal of material may reduce the localized site but the overall concentration of nitrogen is lower and the superconducting properties approach the clean limit.

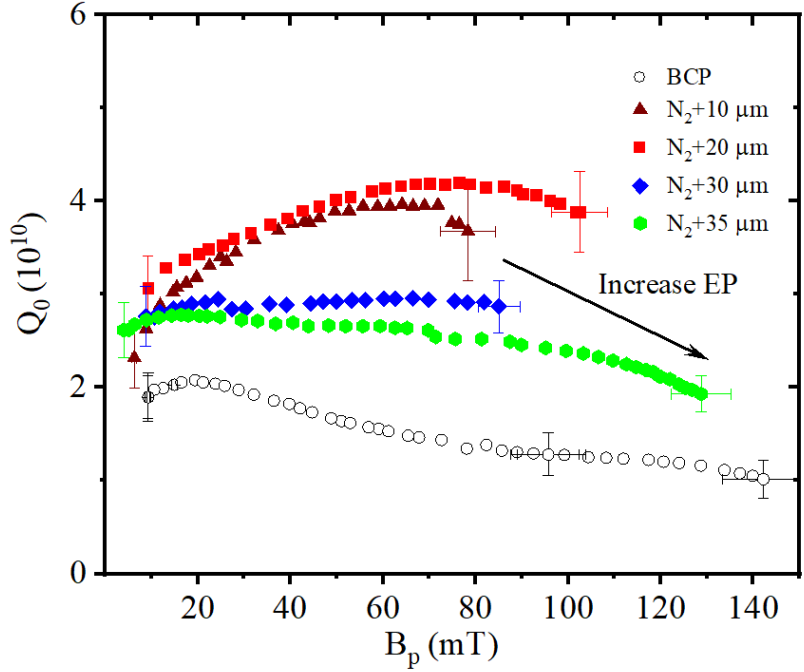

**Figure 8.** The effect of successive EP on a single cell cavity ($B_p/E_{acc} = 4.23\ mT/(MV/m)$) after nitrogen doping [53]. The accelerating gradient increase with increasing the EP removal, while decreasing the overall quality factor.

## IV. Industrialization of Nitrogen Doping

The initial research and development on nitrogen doping were limited to Fermi Lab, Jefferson Lab, and Cornell University. The research groups working closely started developing a "optimum" recipe on single-cell cavities by changing the nitrogen diffusion time[†]. Some details of the recipe development are already mentioned in section III, A. Several trials were done on the exposure and annealing time to identify the best recipe that met the LCLS-II cavity performance specifications. A recipe 2N6 (2 minutes nitrogen exposure followed by 6 minutes of annealing in UHV environment) was chosen to be used for the project and technology was transferred to the cavity manufacturers Research Instruments, Germany and Zanon in Italy. The electropolishing process had already been implemented by the cavity vendors for E-XFEL process but it was refined by Jefferson Lab, mainly cavity temperature, acid flow rate and cathode optimization.

The niobium for cavity fabrication was obtained from Tokyo Denkai, Japan and Ningxia, China, with similar material specifications as for E-XFEL project [62]. A total of 373 cavities were ordered and tested at Jefferson Lab and Fermi Lab and showed an excellent performance with average accelerating gradient $22.0 \pm 3.6$ MV/m with quality factor $(3.1\pm0.5)\times10^{10}$ as shown in Fig. 9 [63,64]. The successful production of several hundred cavities with nitrogen doping proved the reproducibility of the process within a short time. Even though the nitrogen doping was successfully applied to several multi-cell cavities in mass production, several recipe modifications were made during the production [64]. Among them are; issues with electropolishing and annealing process of cavities before the nitrogen doping. The electropolishing issue was mitigated by improved cathode design and parameters with the help of Jefferson lab researchers [63].

---

[†] In literature, the recipe is usually defined as xNy, where x (1-30 mins) represents the nitrogen diffusion time and y (0-60 mins) represent the annealing time in minutes in UHV environment.

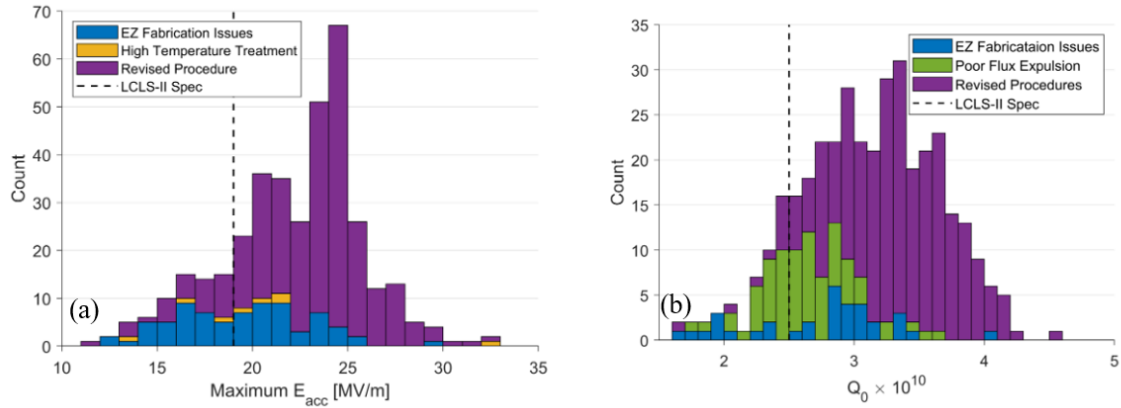

**Figure 9.** Summary of performance of 9-cell LCLS-II cavities (~373 cavities) in vertical test on cavities commercially produced at two different vendor sites, from raw Nb materials from two different vendors. The cavities were doped with 2N6 recipe followed by 5-7 μm EP at vendor sites [63].

To achieve the high quality factor in SRF cavities, one has to minimize the trapped residual magnetic field during the cooldown of SRF cavities. It was found that the high-temperature gradient along the cavity axis is needed during the cooldown when the cavity surface temperature transition from normal conducting to the superconducting state (~9.2 K). This is needed to expel the residual magnetic field from the SRF cavities due to Meissner effect, leading to the lower residual resistance due to trapped magnetic field [65,66,67,68]. It was found that doped cavities are more vulnerable to flux pinning, therefore extra care is needed to minimize any intrinsic source of pinning centers. During the production of LCLS-II cavities, several cavities' quality factor was limited by the poor flux expulsion. Even though the material specification was the same, the performance of the cavities made from the two vendors showed different flux expulsion characteristics. In the middle of the production, additional R&D was done to maximize flux expulsion by increasing the annealing temperature before the nitrogen injection process. The cavities annealed at a higher temperature (900-975 ℃) showed much better flux expulsion compared to cavities annealed at 800 ℃ [64, 69]. The study also showed that the grain size may also play a significant role in flux expulsion [69].

The rapid progress on the development of SRF cavities with nitrogen doping was mainly due to the requirements for the LCLS-II project. As a part of R&D and possible use in future accelerator projects, nitrogen doping work was carried out in other parts of the world, KEK in Japan [33, 70], IHEP [71] and PKU in China [72,73]. Currently, the nitrogen doping technology is considered to be used in the Shanghai Coherent Light Facility (SHINE) linac similar to the LCLS-II project.

# V. Towards High $Q_0$ and High Gradient

The nitrogen doping provided an avenue for the operations of the future accelerators with reduced power consumption due to cavities with high quality factors, particularly CW accelerators which are operated with an accelerating gradient of ≤ 25 MV/m. Coincidently, most of the nitrogen doped cavities are limited to an accelerating gradient ≤ 25 MV/m, significantly lower than that without doping. For example, the average accelerating gradient of 743 cavities used in XFEL projects was 31.4 ± 6.8 MV/m [74,75] with conventional (pre doping era) cavity processing recipe. During the LCLS-II cavity production, the average accelerating gradient of ~373 cavities was 22.0±3.6 MV/m [64]. This shows ~ 30% reduction on an achievable maximum accelerating gradient. However, the higher $Q_0$ at accelerating gradient $E_{acc}$ ~ 16 MV/m, clearly shows the benefit of using doped cavities in CW machine. Future accelerators such as International Linear Collider (ILC) will require a high gradient but it will benefit from SRF cavities with higher $Q_0$ [76]. Moreover, the proposed upgrade of LCLS-HE program is also increasing its operational specification to a $Q_0$ of >2.7×10$^{10}$ at 20.8 MV/m, 2 K, requiring further R&D studies.

## A. High-Temperature Nitrogen Doping

One of the routes to increase the accelerating gradient currently being pursued is to use a modified nitrogen doping recipe by varying the nitrogen diffusion and annealing time. Once again LCLS-HE project-driven R&D is being conducted independently at Jefferson Lab and Fermi Lab pursuing different doping recipes. Jefferson Lab is exploring the doping recipe where the nitrogen is introduced in the furnace for a short time (2-3 minutes) followed by a longer annealing time (~60 mins). The hypothesis behind this recipe is that the nitrogen will diffuse deeper into the bulk with uniform concentration, leading to a high accelerating gradient. Recipe applied to single-cell cavities reached ~35 MV/m, however with the high field Q-slope [77]. Further optimization on doping infrastructure with niobium foil caps installed on cavities opening, smoother pre-doping surface, and colder post doping electropolishing led to initial results on 9-cell cavities of an accelerating gradient closer to 30 MV/m [64, 69].

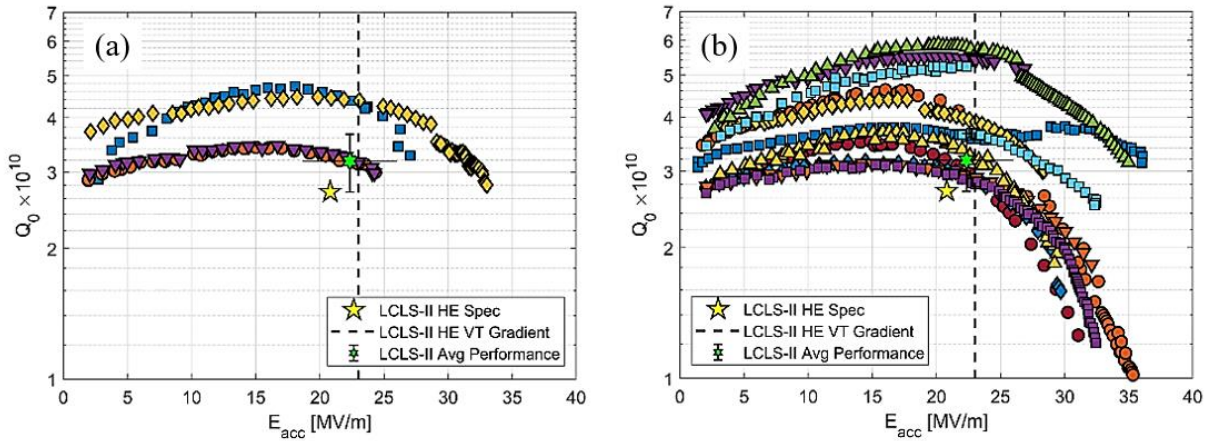

**Figure 10.** $Q_0(E_{acc})$ of single-cell cavities with a modified recipe (a) 2N0 and (b) 3N60, to achieve high $Q_0$, high gradient SRF cavities aim for LCLS-II-HE project [63].

Fermi Lab is exploring a recipe that includes the short nitrogen diffusion time (~2 mins) with no post-annealing time. The preliminary results on a single cell showed the accelerating gradient above 25 MV/m with high field $Q$-slope, similar to that which has been observed with Jefferson Lab 3N60 recipe, as shown in Fig. 10. With available 9-cell cavity test results, using 2N0 recipe, there is a small increase in accelerating gradient (~ 3 MV/m) over the earlier 2N6 recipe [78]. Further R&D is planned to increase the accelerating with high $Q_0$ in doped cavities for the realization of the LCLS high energy upgrade [79].

### B. Recipe Development at Low Temperature

An alternative to the nitrogen doping at high temperature followed by the electropolishing to control the optimal nitrogen concentration within the rf penetration depth was realized by adding nitrogen to the low-temperature baking used to mitigate the high-field Q-slope [11]. In these new so-called nitrogen "infusion" processing recipes, the cavities are heat-treated at 800 °C for ~3 hours then the furnace is allowed to cool to 120-300 °C. The nitrogen is introduced into the furnace at a partial pressure of ~ 25 mTorr for several hours (24-96 hours) in the temperature range of 120-200 °C. This process has shown an improvement in $Q_0$ over the baseline measurements, without the need for post-furnace electropolishing [11, 12, 13, 80, 81,82,83,84].

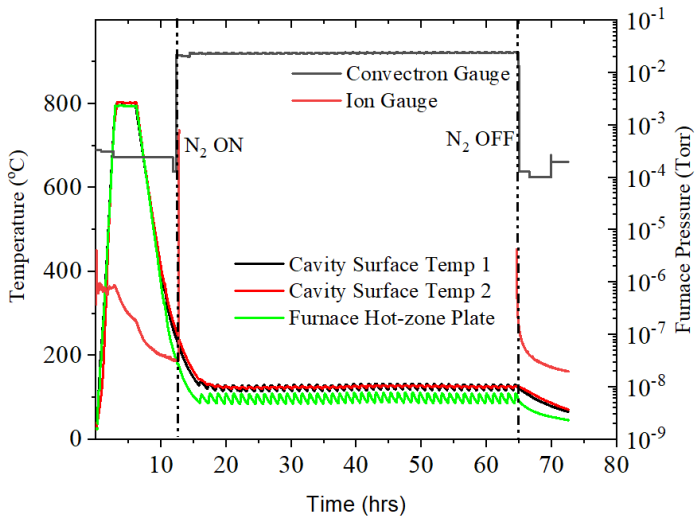

**Figure 11.** Typical temperature and pressure profile during nitrogen infusion run at Jefferson Lab. The nitrogen was injected into the furnace at elevated temperatures [85].

In particular, at Jefferson Lab, a benefit to the cavity performance similar to that achieved by nitrogen doping followed by EP was observed when the nitrogen gas was introduced into the furnace at elevated temperature (~ 250-300 °C) and let the cavity cool down to lower temperature (120-200 °C) and hold at that temperature for an extended period of time (24 - 48 hours) [13, 85]. Typical temperature and pressure profile during nitrogen infusion at Jefferson Lab are shown in Fig. 11. Even though the diffusion of the nitrogen into the bulk of the SRF cavity is limited at these low temperatures (120-200 °C), the introduction of nitrogen is sufficient to modify the cavity surface within the rf penetration depth as seen from cavities' test results, which are similar to those previously reported for high-temperature nitrogen-doped cavities. The nitridation and diffusion of nitrogen into the bulk are expected when the Nb surface is free of $Nb_2O_5$ which occurs >300 °C [86,87], but the infusion was done when the cavity was fully annealed at 800 °C/3h and cooled down to the desired temperature under UHV conditions before injecting nitrogen into the furnace. While post-doping electropolishing is required to remove coarse nitrides from the surfaces of high-temperature nitrogen-doped cavities, no further processing is required for the low-temperature "infusion" recipe showing a clear benefit in reducing processing steps as well as keeping higher gradient with high $Q_0$ values. In some cases, the increase in accelerating gradient over the baseline was also reported, although the actual mechanism isn't fully understood [11, 88]. One of the critical requirement for these low-temperature recipes is that special cleaning and preparation is needed before loading the cavity into the furnace. The cavity is high pressure rinsed (HPR) and then dried in an ISO 4/5 cleanroom. While in the cleanroom, special caps made from niobium foils were placed to cover the cavity flange openings. The cavity was then transported to the furnace inside a clean, sealed plastic bag. To this date, the recipe was successfully applied to single-cell cavities by several research groups, but no clear advantage was demonstrated in multi-cell cavities. The summary of the single-cell cavity results is shown in Fig. 12.

In the quest for high gradient, high quality factor SRF cavities, a two-step low-temperature baking process showed an increase in accelerating gradient as well as quality factor well above 40 MV/m [89,90,91] in single-cell cavities. Furthermore, the medium temperature baking also showed the improved quality factor with a high accelerating gradient [92]. Future accelerating projects need continuous R&D towards high gradient, high quality factor of SRF cavities that can be reliably produced in large numbers.

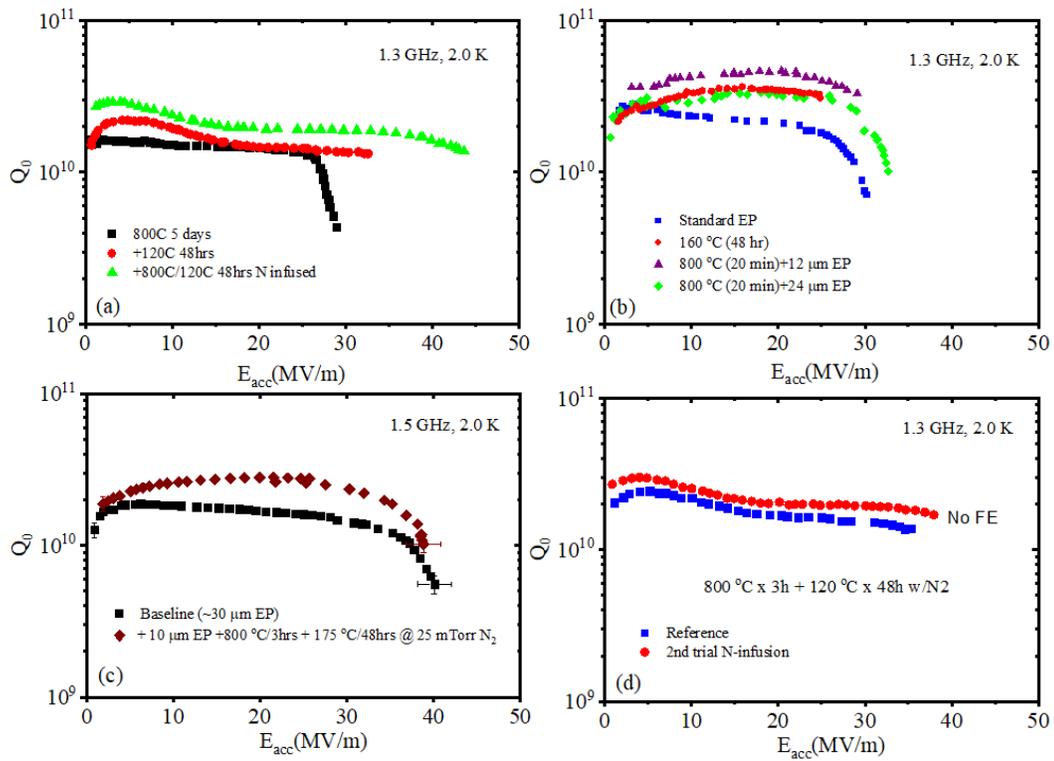

**Figure 12.** The quality factor of single-cell SRF cavities with nitrogen infusion, at (a) Fermi Lab [11], (b) Cornell University [12], (c) Jefferson Lab [13] and (d) KEK [83].

# VI. Sample Coupon Studies on Doping and Infusion

Sample coupons treated along with SRF cavities provide very useful information about the mechanism behind the rf performance of nitrogen-doped and infused SRF cavities. Several surface sensitive techniques are: Scanning electron microscope, Electron diffraction X-ray, Transmission electron microscopy, secondary ion mass spectroscopy, X-ray photoelectron spectroscopy, Magnetometry, and Tunneling.

## A. Surface Morphology

Surface imaging with SEMs along with EDX allowed looking into the surface of niobium after the nitrogen doping. Almost all high-temperature doping recipes produce triangular or star-like shaped structures on the surfaces of niobium (Fig. 13). These structures are identified as normal conducting nitride, mainly β-$Nb_2N$ [46] consistent with the phase diagram shown in Fig. 5. These normal conducting nitrides are removed by electropolishing to achieve the high $Q_0$ in SRF cavities. It was found that the density and size of the normal conducting niobium nitrides depend on the temperature, duration of nitrogen doping, and grain orientation of niobium [93].

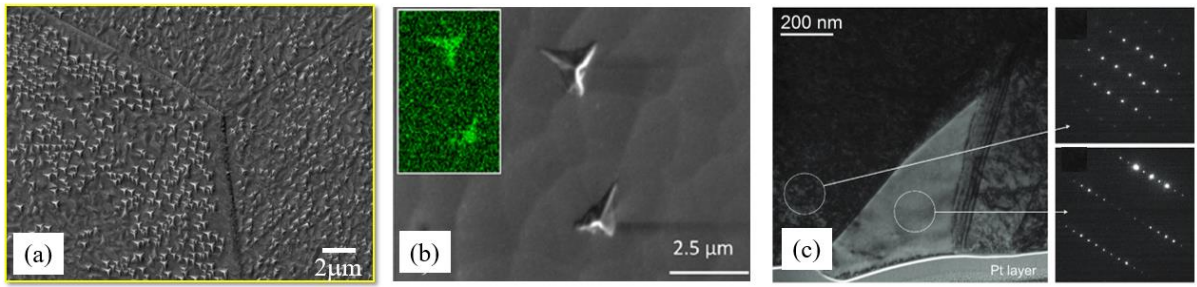

**Figure 13**. (a) SEM image of Nb surface after nitrogen doped with 2N6 recipe at 800 °C [55], (b) SEM and EDX image (inset) elemental map for $NK_\alpha$ of nitrogen doped (20N30) sample [47] and (c) TEM image of nitrogen doped (20N0) Nb sample with nano electron diffraction image of Nb, [113] zone axis (upper) and of $Nb_2N$ close to [310] zone axis (lower) [46].

## B. Secondary Ion Mass Spectroscopy (SIMS)

SIMS has been an excellent tool to measure elemental concentration on the surface of the nitrogen-doped Nb samples. Ions beams, primarily $^{16}O_2^+$, $^{16}O^-$, $^{133}Cs^+$, or $B_n^{m+}$ (n=1-5, m=1,2) are used to bombard the sample surface at an energy that is enough to eject secondary ions from the surface of the Nb. The ejected ions are detected and analyzed by the mass spectrometers to infer the composition of the sample [94]. The concentration of nitrogen as a function of depth profiles and other impurities such as oxygen, hydrogen, carbon dioxides, and carbon monoxides are typically measured during the SIMS study depending on the detection limit of the instrumentation. It is to be remembered that the quality factor of SRF cavities depends on the elemental compositions (impurities) within the rf penetration depth. The additional rf losses can be anticipated due to the large segregation of impurities even deeper than the rf penetration depth since those sites act as effective pinning centers for the residual magnetic field and thereby contributing to the vortex related rf loss [95]. The concentration profiles measured on nitrogen-doped Nb samples typically follow the simulation results using Fick's laws [44]. As expected, the variation on nitrogen concentration is observed depending on the nitrogen partial pressure [45], the temperature of diffusion [52], and the duration of nitrogen doping [63] as shown in Fig. 14.

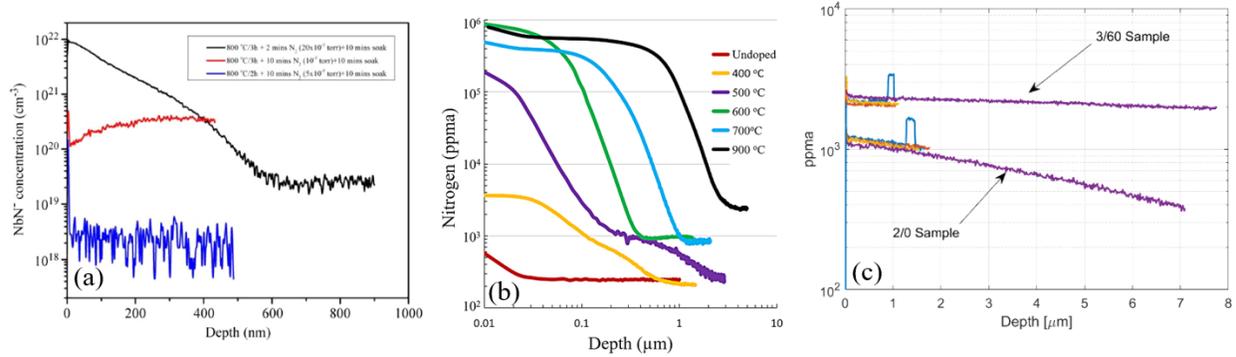

**Figure 14.** Diffusion depth profile of nitrogen to the (a) partial pressure of nitrogen [45], (b) temperature of nitrogen exposure [52], and (c) duration of nitrogen exposure [63].

SIMS measurements were also carried out samples treated in nitrogen environments at low temperatures and also with titanium doped samples [96]. The diffusion of nitrogen within the rf penetration depth (~50 nm) was observed when the baking temperature was kept 150-200 ºC [11, 13, 47, 97]. SIMS measurements shown the increased concentration of carbon and oxygen in nitrogen infused niobium sample [98]. Higher concentrations of oxygen were observed as a result of heat treatment of Nb samples above 120 ºC even in the UHV environment as oxygen diffuse much deeper than nitrogen or carbon in niobium [12, 55, 70, 96, 99].

C. X-ray photoelectron spectroscopy (XPS)

XPS studies are done to identity the near-surface composition on SRF niobium. Historically there have been several XPS studies done to understand the loss mechanism of SRF cavities on EP, BCP and Low temperature baked SRF niobium [100,101]. XPS studies on those treatments reveal a complex oxide substructure ($Nb_2O_5$, $NbO$, $NbO_2$, $NbN_{1-x}O_x$) within the rf penetration depth (~10 nm), as shown in Fig. 15, which can be responsible for the rf performance of SRF cavities [13, 47, 55, 71].

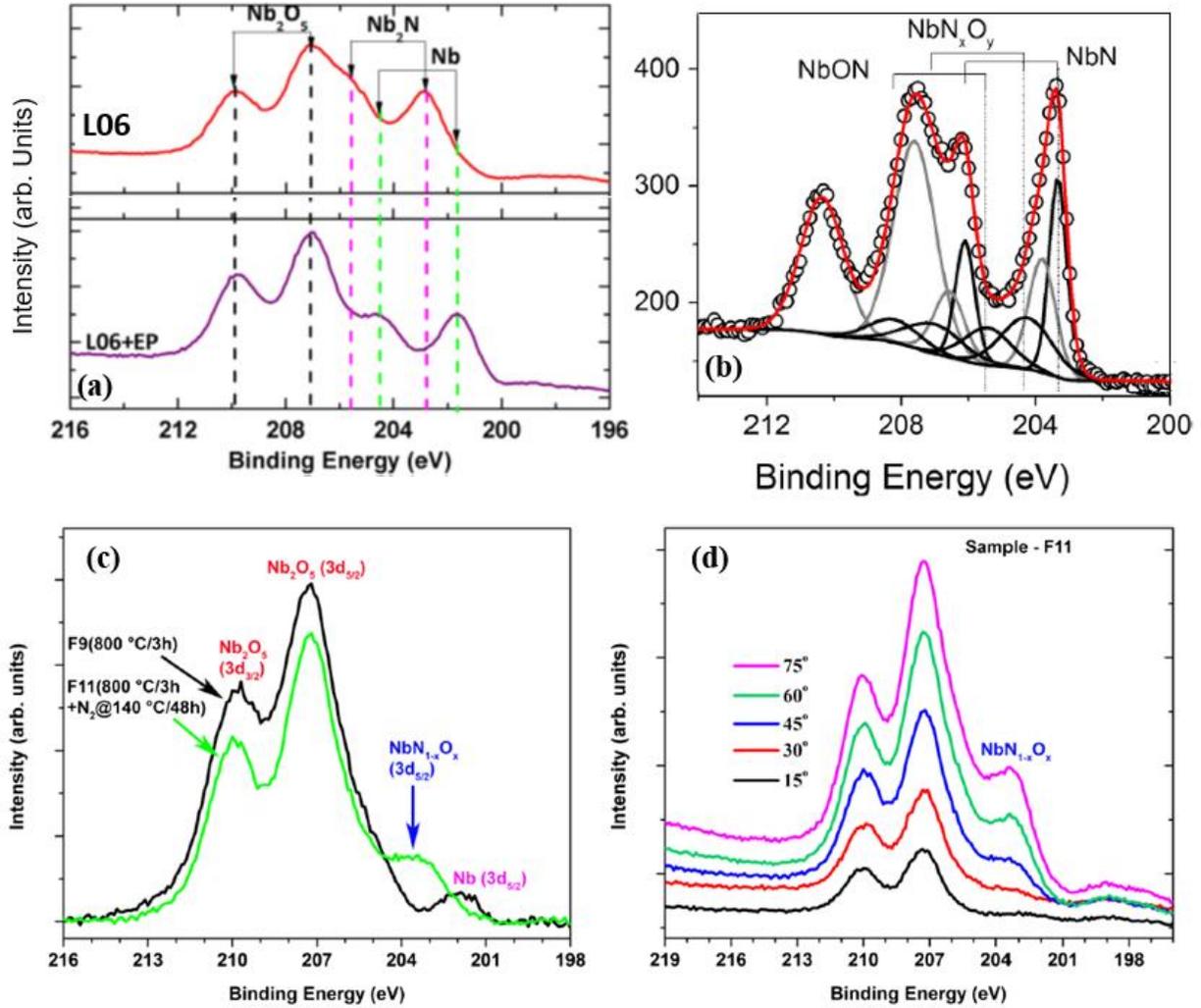

**Figure 15**. (a) Nb 3d XPS spectra (takeoff angle of 75°) of N-doped (2N6) at 800 °C Nb samples before and after the ~6 μm surface removal with EP [55]. (b) Deconvoluted Nb 3d lines measured for Nb samples after nitridation at 800 °C [47]. (c) Background subtracted XPS spectra on sample F9 (800 °C/3 hrs, no doping) and F11 (800 °C=3 hrs +140 °C/48 hrs at 25 mTorr $N_2$) at 45° takeoff angle. And, (d) Angle-resolved XPS on sample F11 (low-temperature nitrogen infusion sample) shows the complex $NbN_{1-x}O_x$ within the first 10 nm of the Nb surface [13].

### D. DC Magnetization and AC Susceptibility

The superconducting properties such as the transition temperature and critical fields can be extracted from the dc magnetization and susceptibility measurements. In particular, dc magnetization can be used to estimate the field of first flux penetration ($H_{ffp}$), which is believed to be one of the causes of cavity quench. The measurement showed that $H_{ffp}$ is lowered by ~ 15% due to the nitrogen doping (Fig. 16 (a)) [102, 103,104] with no significant change in the bulk superconducting properties, confirming that the doping volume is smaller compared to the sample size.

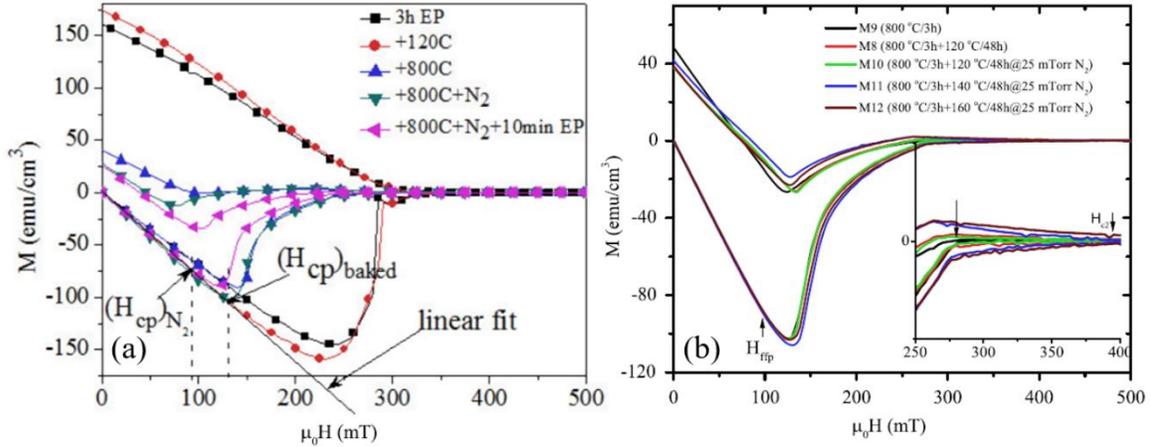

**Figure 16**. Isothermal dc magnetic hysteresis of samples after (a) several surface treatments including nitrogen doping and followed by EP [103]. (b) Sample with nitrogen infusion at low temperature [13]. About a 15% reduction in first flux penetration was observed after nitrogen-doped samples, whereas no significant change in $H_{ffp}$ was observed in low-temperature nitrogen-infused samples.

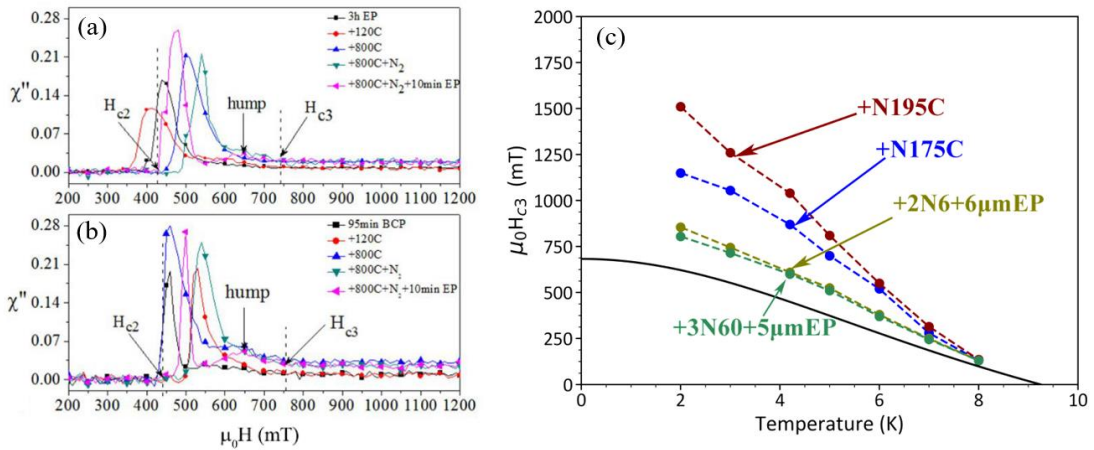

**Figure 17**. Imaginary part, $\chi''$, of dc magnetic field-swept-mode ac susceptibilities of a nitrogen-doped sample with initial surface (a) EP, (b) BCP, followed by EP [103]. And (c) the surface critical field of clean Nb (solid line), nitrogen-doped Nb followed by EP and low-temperature nitrogen-infused Nb samples [105].

The ac susceptibility measurement technique has been used to study the pinning and vortex motion in many superconductors. The real part of the susceptibility is the result of the diamagnetic shielding of the magnetic field whereas the imaginary part gives information about the loss mechanism. The external magnetic field dependence of the ac susceptibility can be used to extract the surface critical fields $H_{c1}$ and $H_{c2}$ as shown in Fig. 17. In some geometrical conditions, when the external magnetic field is parallel to the sample surface, the nucleation of the superconducting phase in a thin surface sheath was first discovered by Saint-James and de Gennes [106] up to the surface critical field $H_{c3}$. The ratio $H_{c3}/H_{c2}$ was estimated to be ~1.7 using Ginsburg-Landau theory. It has been previously reported that the experimentally measured ratio is higher than the theoretical number for several Nb samples [107,108] and the ratio increases with decreasing the electronic mean free path. The ratio for nitrogen-doped niobium was found to be closer to the theoretical limit [101,103], even though, the electronic mean free path was reduced as a result of nitrogen diffusion into the Nb surface.

### E. Tunneling Measurements

The BCS surface resistance depends on many superconducting parameters and among them are the superconducting gap and density of states (DOS) of the quasi-particles at the Fermi level. The tunneling spectroscopy such as point contact tunneling (PCT) and scanning tunneling microscope (STM) is particularly useful in determining the local DOS of quasi-particles and superconducting gap as well as imaging vortices. PCT measurements done on nitrogen-doped Nb showed a more homogeneous distribution of the superconducting gap compared to non- doped samples and also favorable surface oxidation states on the surface [109]. The oxide structure is related to rf losses related to two-level systems [110]. Recent STM measurements on cut out samples from nitrogen-doped cavities showed the homogeneity of the superconducting gap with broadening of DOS [111] predicted for dirty superconductors [112]. Furthermore, the vortex imaging with STM showed a slightly reduced but homogenous superconducting gap and lower coherence length compared to the traditionally prepared niobium cavities.

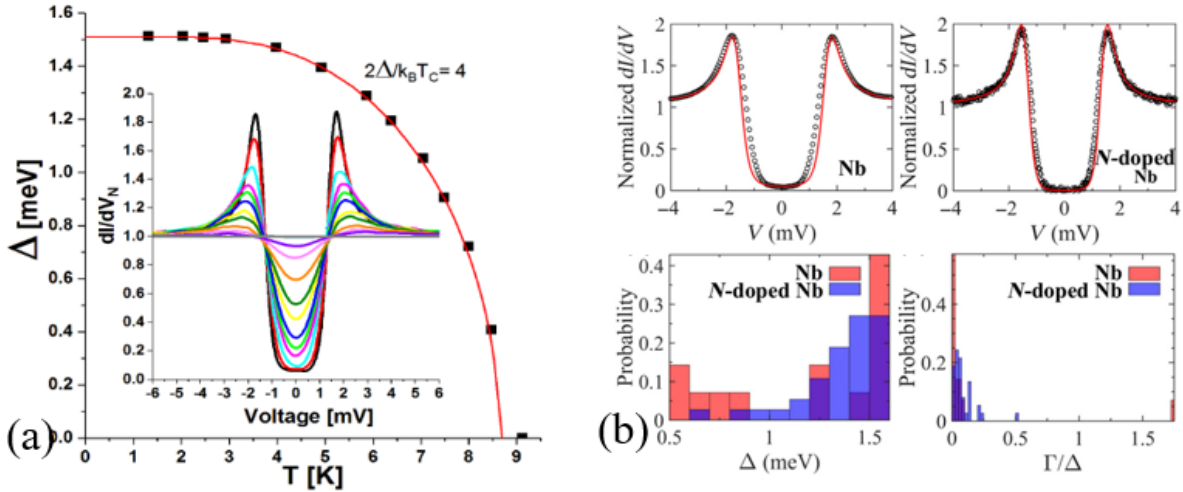

**Figure 18.** (a) Temperature dependence of Δ typical tunnel junction measured using point contact tunneling (PCT) shown in the insert [109]. (b) Scanning tunneling spectrum (dots) and BCS-Dynes fit (red line) acquired on Nb and N-doped Nb cavity cutout samples, respectively, at T=1.5 K [111].

## VII. Theoretical Models

As mentioned earlier, the increase in quality factor can be realized by minimizing the surface resistance, either residual or BCS, or both. Statistically, it was found that the residual resistance of doped SRF cavities is lower than the non-doped counterpart when rf tested in the low residual magnetic field environment. It is to be noted that the residual magnetic field has a severe effect on doped cavities compared to non-doped cavities [68,113,114,115]. Historically, it is believed that the hydrogen in niobium is responsible for higher residual resistance due to the precipitation of hydrogen forming normal conducting hydrides within the rf penetration depth [116,117,118,119]. Some improvement in quality factor at the low accelerating gradient and a significant improvement in quality factor with increased accelerating gradient was observed when cavities were baked at low temperature. Even though no unanimous theory or model exists at this time, evidence shows that the impurities (O and H), as well as their interactions with vacancies, play a role in minimizing rf losses in SRF cavities by the low-temperature baking [120,121,122,123,124]. The presence of nitrogen (or even titanium) traps the hydrogen, preventing the formation of lossy hydrides probably leading to the lower residual resistance in doped SRF cavities [125,126,127, 128, 129,130].

One of the peculiar observations in the doped cavity is the increase in quality factor as a function of the microwave field tangential to the inner surface of the cavity. This means that the surface resistance decreases with the rf field, contrary to the perception that the microwave field suppressed the superconductivity leading to the decrease in quality factor $Q_0$ with accelerating gradient $E_{acc} \propto H_{rf}$ [131]. The quality factor of doped SRF cavities increases with the increase in rf field up to $H_{rf} \sim 100$ mT and is then limited by either quenching or by the high field $Q$-slope, typically observed in non-doped BCP or EP cavities.

The original BCS theory was developed without taking into account the amplitude of the rf field and the theory explains well the temperature dependence of surface resistance at low rf field [132,133]. The theory was extended to explain the high field $Q$-slope in SRF cavities, which may be related to the non-linear BCS resistance [134,135]. Soon after the first demonstration of $Q$-rise phenomena after the titanium doping in SRF cavities [9], the BCS theory was extended taking into account the motion Cooper pairs in the presence of rf field [136]. The calculation reproduced the rf field dependence surface resistance of titanium doped cavity and similarly, nitrogen-doped cavities [137].

Gurevich [112] proposed that microwave suppression of surface resistance comes from the current-induced broadening of the quasiparticle density of states. The BCS surface resistance takes a logarithmic dependence on the amplitude of the rf field in the dirty limit, as experimentally observed in Ti [138] and nitrogen-doped cavities [53,139,140,141]. The microwave suppression of surface resistance was also previously observed in thin films and explained due to the non-equilibrium quasi-particle distribution function leading to the enhancement in supercurrent [142]. A theoretical model in which the surface resistance of a superconductor coated with a thin normal metal was recently presented and showed that the $R_s(B_p)$ behavior observed in SRF cavities following different surface preparations can be explained by changes in the thickness of the normal layer and of the interface boundary resistance [143]. A recent theoretical model proposed by Gurevich extends the zero-field BCS surface resistance to high rf fields, in the dirty limit [144]. Such a model calculates $R_s(H)$ from the nonlinear quasiparticle conductivity $\sigma_1(H)$, which requires knowledge of the quasiparticles' distribution function. The calculation was done for two cases, one which assumes the equilibrium Fermi-Dirac distribution function and one for a non-equilibrium frozen density of quasiparticles. A non-equilibrium distribution function is appropriate when the rf period is shorter than the quasiparticles' relaxation time. $R_s(H)$ was calculated numerically for these two cases and it depends on a single parameter, $\alpha$, which is related to the heat transfer across the cavity wall, the Nb-He interface, and between quasiparticles and phonons. The calculation based on this model qualitatively reproduces the microwave surface resistance in cavities with low-temperature nitrogen infusion [13].

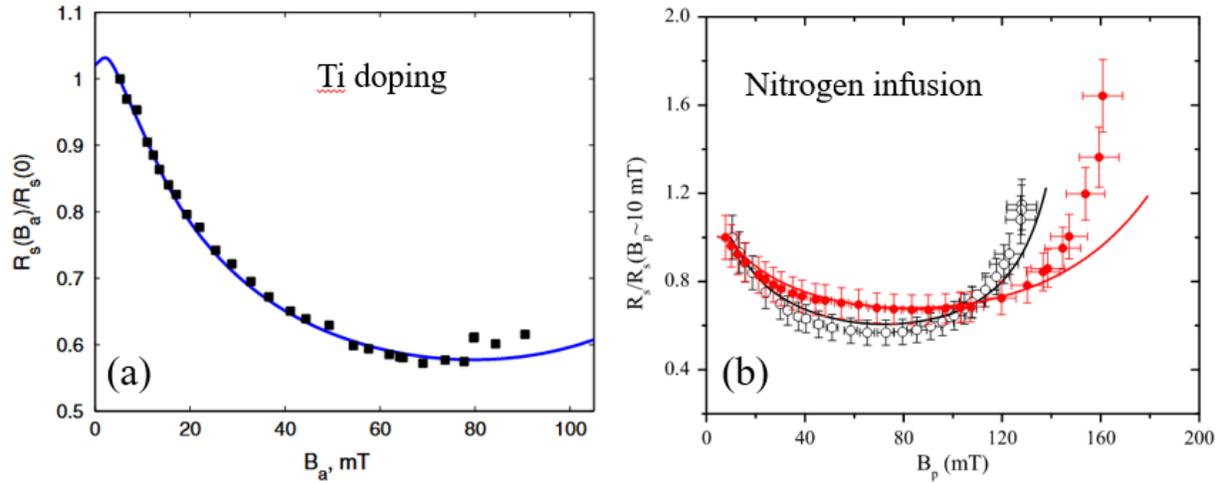

**Figure 19**. Normalized surface resistance as a function of peak surface magnetic field on cavity surface (a) Ti-doped cavity [112] and (b) nitrogen-infused SRF cavities [13], the solid lines are calculated using the model described in refs. 112 and 144.

Several other models were reported to understand the microwave resistance in SRF cavities extending the theory to clean, dirty, equilibrium, and non- equilibrium superconductors [145,146, 147,148]. The nitrogen doping and infusion studies were also extended to cavities with resonant frequency other than 1.3 and 1.5 GHz [149,150], showing that the frequency might play the role in the field dependence of microwave surface resistance [146]. Also, two-fluid models taking into account surface impurities on SRF cavities were extended to fit the field-dependent surface resistance and the model was used to extract the superconducting parameters responsible for the reduced rf loss in nitrogen-doped SRF cavities [151,152].

## VIII. Summary and Future Outlook

Engineering the niobium surface with impurity doping within the rf penetration depth in SRF cavities has been extremely beneficial to achieve high quality factor, potentially cost-saving on current and future SRF based accelerators. SRF cavities with high quality factor ($Q_0 > 3 \times 10^{10}$) at 2.0 K and accelerating gradient ~ 20 MV/m have been successfully produced commercially and are in the process of installation at SLAC for the LCLS-II project. Efforts are being made to extend high quality factors towards higher accelerating gradient with modification of existing recipes as well as exploring new recipes. High $Q_0$ and high gradient SRF cavities would be of great interest for lowering the cryogenic heat load of future high-energy accelerators.

High quality factor SRF cavities are also drawing interest in several other applications. For example, quantum computing with microwave resonators [153,154,155] and axion dark matter research [156]. An extremely high quality factor with a decay time of the order of few seconds was measured in SRF cavities in the quantum temperature limit [157]. In addition to Nb cavities, alternative materials $Nb_3Sn$ are being explored to achieve high quality factor at higher temperature, however, so far the accelerating gradient is limited to below 20 MV/m [158,159,160]. $Nb_3Sn$ cavities are fabricated using thermal diffusion of Sn on Nb cavities and also have the potential for industrial applications [161,162,163]. For high accelerating gradient cavities, multilayers of superconductors are being considered [164,165] as a solution to the low $H_{c1}$ in strong type-II superconductors, such as $Nb_3Sn$.

## IX. Acknowledgments

We would like to acknowledge Dr. G. Ciovati, Dr. R. Geng, Dr. A. Palczewski, and Dr. Robert Rimmer from Jefferson Lab for helpful discussions. We would like to acknowledge Dr. S. Balachandran from Applied Superconductivity Center, NHMFL with fruitful discussions. This manuscript has been authored by Jefferson Science Associates, LLC under U.S. DOE Contract No. DE-AC05-06OR23177.